\documentclass[%
preprint,
amsmath,amssymb,
aps,
]{revtex4-2}

\usepackage{graphicx}
\usepackage{bm}

\renewcommand{\thesection}{\arabic{section}}
\renewcommand{\thesubsection}{\thesection.\arabic{subsection}}
\renewcommand{\thesubsubsection}{\thesubsection.\arabic{subsubsection}}

\makeatletter
\renewcommand{\p@subsection}{}
\renewcommand{\p@subsubsection}{}
\makeatother

\begin{document}

\title{Generalized model of incipient plasticity with parametric variations}

\author{Sweta Kumari}
\author{Aditya Vardhan Mishra}
\author{Amlan Dutta}
\email{amlan.dutta@metal.iitkgp.ac.in}
\affiliation{Department of Metallurgical and Materials Engineering, Indian Institute of Technology Kharagpur, West Bengal 721302, India}

\date{\today}

\begin{abstract}
	Incipient plasticity is typically associated with thermally activated events like the nucleation of dislocations in crystalline solids and the activation of shear transformation zones in metallic glasses. A widely employed method of estimating the activation parameters of such mechanisms involves analyzing the statistical distribution of critical loads obtained through a series of repeated measurements. However, the conventional mathematical approach assumes the activation parameters to remain fixed during the sequence of measurements. The present study critically examines this premise and presents a generalized statistical model that allows the statistical variations of activation parameters. Using a simple Monte Carlo scheme, it is demonstrated that even small fluctuations of activation parameters can significantly affect the statistical distribution of measured critical loads. The Monte Carlo calculations, along with atomistic simulations, further show that imposing the assumption of rigidly fixed parameters on an activated process with parametric fluctuations can lead to severe underestimation of the activation parameters. As many experimental studies have consistently reported perplexingly small activation volumes estimated using the conventional statistical approach, we propose that our findings can offer a fresh perspective on this longstanding issue.
\end{abstract}

\maketitle

\section{\label{sec:level1}Introduction}

Experiments like nano-indentation and tensile or compressive loading of solids exhibit abrupt transitions from elastic to plastic modes of deformation in the form of pop-in events and yield points \cite{pohl2019pop,jiapeng2017,ohmura2021pop,huang2009,zhang2019microplasticity,chen2016elastic,sedore2019comparison}. Such transitions are typically facilitated by load-driven activated processes like nucleation of dislocation loops \cite{mojumder2020,zhao2018,lee2020situ}, depinning of line defects \cite{avalanches,friedman2012}, formation of shear-transformation zones in metallic glasses \cite{avila2019shear,2015strain}, etc. The onset of plasticity is typically determined by a combination of temperature and rate of deformation, implying that the underlying process involves load-driven, thermally activated events. Temperature provides the thermal assistance required by the system to traverse along the minimum energy path, while the applied load effectively reduces the activation barrier along the transition pathway.\\

In any generic load-driven process, the activation volume is a fundamental parameter associated with the kinetics. It represents the rate at which the activation barrier reduces with increasing applied load. Besides, this key parameter also renders a rough estimate of the number of atoms involved in the process and plays a broader role in offering insights into the underlying mechanism. While its direct measurement is straightforward in atomistic computations with methods like the nudged elastic band (NEB) \cite{chen2019atomistic,geslin2017,zhu2007interfacial,kumari2021nucleation}, umbrella sampling \cite{ryu2011,ryu2011entropic}, finite temperature string \cite{weinan2005finite}, etc., experimental methods resort to mathematical models linking the activation volume to other measurable quantities. For instance, the activation volume of solid-state diffusion has been estimated by measuring diffusivities under varying pressure \cite{aziz2006}. In metallic glasses, the activation volume for STZ activities is indirectly determined by measuring the hardness and strain-rate sensitivity \cite{sahu2019mechanism,kramer2018activation}. Similarly, Kato \textit{et al.} ~\cite{kato2009thermally} has demonstrated a line tension model that can be used to estimate the activation volume of dislocation depinning at grain boundaries using experimentally gathered flow stress data at various temperatures.\\

In 2004, Schuh \textit{et al.} \cite{schuh2004} developed an interesting model to extract the activation parameters of a load-driven process by analyzing the statistical distribution of the critical loads of elastic-plastic transition obtained through a series of identical measurements. Although the original method was derived and demonstrated for nano-indentation of crystalline samples, the fundamental idea is generic and easily applicable to various materials and processes. Consequently, this statistical method has witnessed numerous applications in various experimental and \textit{in silico} studies. A specific premise of the original Schuh and
Lund (SL) model \cite{schuh2004} is the assumption of rigid activation parameters of a thermally assisted event. It implies that the statistical approach tacitly assumes that for repetitive measurements of critical load in an experiment, each measurement is either the outcome of the same transition pathway with identical activation parameters or the parametric variation from one measurement to another is small enough to affect the calculations significantly. A follow-up study \cite{mason2006} considered the possible effects of surface defects that could affect the resulting statistics of pop-in loads measured during the nano-indentations. However, such factors are assumed to have only secondary effects on the statistical distribution.\\

In the present study, we revisit the fundamental premise of the statistical method of determining activation energy and volume. It involves examining the assumption that an identical event with fixed activation parameters occurs every time the critical load is recorded during the critical load measurements. Furthermore, we present a generalized version of the statistical model, which allows the activation parameters to fluctuate during repeated measurements. To explore the ramifications of such parametric variations, a Monte Carlo scheme is used to numerically implement the model and compare the results to those obtained from the conventional rigid-parameter model. This aspect is explored further through molecular dynamics simulations of tensile and compressive deformation of iron nanopillars. As discussed subsequently, the simulation outcomes show that the activation volumes computed through the rigid-parameter Schuh-Lund model appear to be incommensurate with the yielding mechanisms shown by the simulations. We point out that most of the experiments resorting to the rigid-parameter model also report activation volumes that are much smaller than the expectations from the possible mechanisms of incipient plasticity. As the generalized statistical model exhibits a significant impact of relatively small parametric variance on the distribution of measured critical loads, it hints at the possibility of underestimated activation volumes in the reported studies on account of the presumption of rigid activation parameters in the conventional SL model.

\section{\label{sec:level1}Rigid-Parameter Model}

The statistical model developed by Schuh and
Lund \cite{schuh2004} offers the advantages of simplicity and an exact solution. The model is based on the realization that for a given loading rate, the rate of increase of the cumulative probability (\emph{F}) of a transition event is given by the product of the survival probability (1-\emph{F}) and transition rate ($k$). Assuming that the activation-free energy reduces linearly with increasing load over the range of interest, we have the transition rate,

\begin{equation}
	k = \nu_{0}e^{- \Delta G/kT}, 
\end{equation}

\noindent with \(\nu_{0}\) as the attempt frequency. The free energy barrier, \(\Delta G = G_{0} - v^{\ast}\sigma\) depends on the activation volume, \(v^{\ast}\), and the relevant stress component, \(\sigma\). Solving the resulting differential equation for the cumulative probability renders the solution,

\begin{equation}	
	F(\sigma) = 1 - \exp\left\lbrack - \frac{\eta kT}{\dot{\sigma}v^{\ast}}\exp\left( \frac{\sigma v^{\ast}}{kT} \right) \right\rbrack,
\end{equation}

\noindent which can rewritten as,

\begin{equation}
	\ln{\left\lbrack \ln(1 - F)^{- 1} \right\rbrack\  = \ \ln{\frac{\eta kT}{\dot{\sigma}v^{\ast}} + \frac{\sigma v^{\ast}}{kT}}},
\end{equation}

\noindent where \(\eta = \nu_{0}e^{- G_{0}/kT}\) and \(\dot{\sigma} = d\sigma/dt\).\\

The basic essence of the statistical model can be captured through a simple scheme of Monte Carlo calculations. We assume that the load, \(\sigma\), increases iteratively in small steps of 0.01 per iteration, starting with \(\sigma = 0\). In each iteration, corresponding to a time duration of unity, the barrier, \(\Delta G\), is evaluated, and a uniformly distributed random number is generated. If this number is found to be less than \(e^{- \Delta G/kT}\), the activated event is considered to have taken place, and the iterations are terminated. In the context of Eq.~(3), this scheme simulates a transition event with \(\nu_{0}\) = 1 and \(\dot{\sigma} = 0.01\) in arbitrary units. We perform calculations with the parameters, \(G_{0} = 1\), and \emph{kT} = 0.05. For each selected value of the activation volume, the Monte Carlo calculations are executed 500 times with different initializations of the random number generator.\\

Figure~1 displays the variation of reciprocal of the survival probability $(1-F)$ scaled to a double logarithmic form. For three different activation volumes, \(v^{\ast}\) = 0.8, 1.0, and 1.2, the plots exhibit linear behavior as predicted by Eq.~(3). The excellent agreement between the distributions given by the Monte Carlo calculations and those obtained independently from Eq.~(3) verifies the mathematical correctness of the SL model and validates the Monte Carlo scheme.

\begin{figure}
	\centerline{\includegraphics*[width=8cm, angle=0]{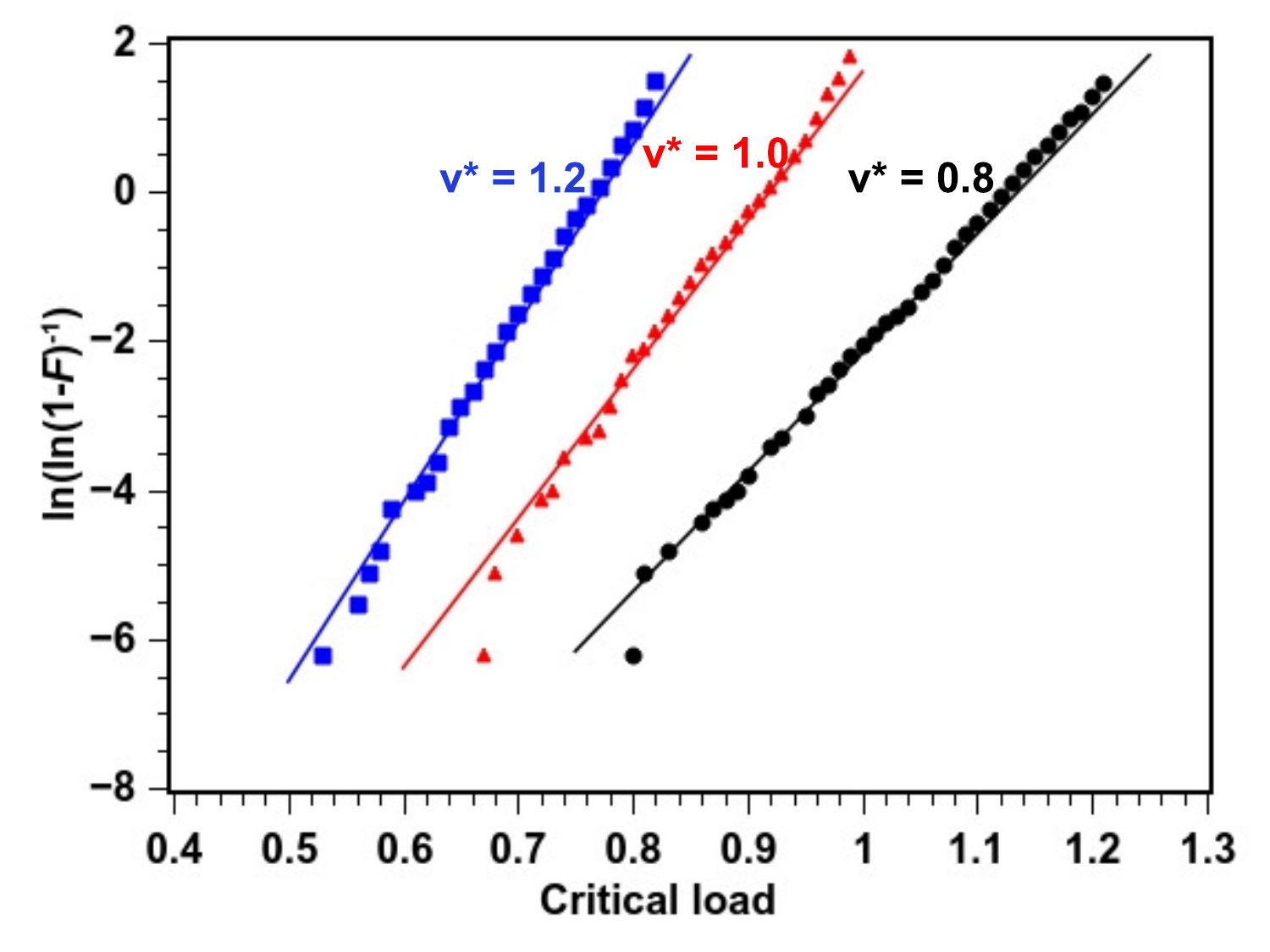}}
	\caption{Variation of the reciprocal of survival probability, $1-F$, shown against the critical load. The values on the ordinate are shown with double logarithmic scaling. The scattered symbols present the outcome of Monte Carlo calculations while the solid lines represent the analytical result of Eq.~(3).}
\end{figure}

\section{\label{sec:level1}Generalization with parametric fluctuations} 

Given the preceding discussion, we can verify whether some dispersion in the activation parameters within narrow ranges can indeed produce an acceptable statistical estimate of their representative central values. To this end, we may revisit the original statistical approach and introduce the statistical variations of activation parameters into the model. Consider a set of experiments involving measurements of the critical load of incipient plasticity at a given temperature and loading rate. Every measurement involves a transition pathway, \emph{i}, which is picked from a pool of multiple possible event-types with probability, \(\rho(G_{0}^{i},\nu_{0}^{i},v_{i}^{\ast}, \sigma)\), with the corresponding activation parameters, \(G_{0}^{i}\), \(\nu_{0}^{i}\), and \(v_{i}^{\ast}\) (Sec.~2). If $dN_i$ is the number of measured load samples in the infinitesimal range of \(\sigma\) to \(\sigma + d\sigma\), corresponding to event-type \emph{i}, we have,

\begin{equation}
	dN_{i} = N_{i}f_{i}(\sigma)d\sigma,
\end{equation}

where $N_{i}$ is the total number of measured critical loads resulting from transition event, \textit{i}. In the above equation, \(f_{i}(\sigma)\) is the probability density of the particular event-type with respect to variation in the measured critical load, \(\sigma\). Summing over all the possible event-types, we obtain,

\begin{equation}
	dN = \sum_{i}^{}{dN}_{i} = \sum_{i}^{}{N_{i}f_{i}(\sigma)d\sigma}. 
\end{equation}

Since the experimental or \emph{in silico} load measurement cannot identify the event type for a particular measurement, direct estimation of the probability distribution, \(f_{i}(\sigma)\), is infeasible. Nonetheless, the measurements still provide us with the overall distribution, \emph{f}(\(\sigma\)), of all the sampled loads collectively and \(dN = Nf(\sigma)d\sigma\). As, \(N_{i} = N\rho(G_{0}^{i},\nu_{0}^{i},v_{i}^{\ast},\sigma)\), Eq.~(5) yields,

\begin{equation}
	f(\sigma)d\sigma = \sum_{i}^{}{\rho(G_{0}^{i},\nu_{0}^{i},v_{i}^{\ast},\ \sigma)f_{i}(\sigma)d\sigma},
\end{equation}

and the cumulative distribution is obtained as,

\begin{equation}
	\Phi(\sigma) = \sum_{i}^{}{\int_{- \infty}^{\sigma}{\rho(G_{0}^{i},\nu_{0}^{i},v_{i}^{\ast},\ \sigma)f_{i}(\sigma)d\sigma}}.
\end{equation}

It must be noted that the distribution,\(\ \Phi(\sigma)\), includes all the possible event-types, \emph{i}, and is fundamentally distinct from the cumulative distribution given in Eq.~(2), which intrinsically assumes a single set of fixed activation parameters. Although Eq.~(7) is derived with the assumption of a discrete set of possible parameters, we can easily generalize it to include a continuous distribution over the parametric space as given below,

\begin{eqnarray}
	\Phi(\sigma) &=& \int_{G_{0}^{\min}}^{G_{0}^{\max}} \int_{\nu_{0}^{\min}}^{\nu_{0}^{\max}} \int_{v_{\min}^{\ast}}^{v_{\max}^{\ast}} \int_{-\infty}^{\sigma} \rho(G_{0},\nu_{0},v^{\ast},\sigma)f(G_{0},\nu_{0},v^{\ast},\sigma) \, d\sigma \, dv^{\ast} \, d\nu_{0} \, dG_{0}.
\end{eqnarray}

Equation~(8) represents an extension of the original statistical scheme in its most generic form, where in contrast to Eq.~(2), the activation parameters, $G_{0}$, $\nu_{0}$, and $v^{\ast}$ are allowed to exhibit statistical fluctuations during repeated measurements. Various assumptions can be made to reduce it to simplified forms further. For instance, if the parametric distribution, $\rho$, is considered to be independent of the measured load, \(\sigma\), Eq.~(8) becomes,

\begin{eqnarray}
	\Phi(\sigma) &=& \int_{G_{0}^{\min}}^{G_{0}^{\max}} \int_{\nu_{0}^{\min}}^{\nu_{0}^{\max}} \int_{v_{\min}^{\ast}}^{v_{\max}^{\ast}} \rho(G_{0},\nu_{0},v^{\ast})\left\{ \int_{- \infty}^{\sigma} f(G_{0},\nu_{0},v^{\ast},\sigma)d\sigma \right\} dv^{\ast} d\nu_{0} dG_{0}.
\end{eqnarray}

Here, it must be pointed out that while Eq.~(9) provides the cumulative distribution over the whole three-dimensional parametric space, Eq.~(2) still represents the partial cumulative distribution for a specific event-type with fixed activation parameters, \(G_{0}^{}\), \(\nu_{0}^{}\), and \(v^{\ast}\). Hence, Eq.~(9) can be rewritten as,

\begin{eqnarray}
	\Phi(\sigma) &=& \int_{G_{0}^{\min}}^{G_{0}^{\max}} \int_{\nu_{0}^{\min}}^{\nu_{0}^{\max}} \int_{v_{\min}^{\ast}}^{v_{\max}^{\ast}} \rho(G_{0},\nu_{0},v^{\ast})\left[1 - \exp\left\{ -\frac{\eta_{0}kT}{\dot{\sigma}v^{\ast}}\exp\left(\frac{\sigma v^{\ast}}{kT}\right)\right\}\right]dv^{\ast} d\nu_{0} dG_{0}.
\end{eqnarray}

It is pertinent to note that the special case of a single event-type with constant activation parameters, \(G_{0}^{c}\), \(\nu_{0}^{c}\), and \(v_{c}^{\ast}\), is mathematically reproduced by setting the parametric distribution in Eq.~(10) through the Dirac-delta functions as, \(\rho(G_{0},\nu_{0},v^{\ast}) = \delta(G_{0} - G_{0}^{c})\delta(\nu_{0} - \nu_{0}^{c})\delta(v^{\ast} - v_{c}^{\ast})\), thereby retrieving the original result of Eq.~(2) as,

\begin{equation}
	\Phi(\sigma) = 1 - exp\left\{ - \frac{\eta_{c}kT}{\dot{\sigma}v_{c}^{\ast}}\exp\left( \frac{\sigma v_{c}^{\ast}}{kT} \right) \right\},  
\end{equation}

with \(\eta_{c} = \nu_{0}^{c}e^{- G_{0}^{c}/kT}\).

\section{\label{sec:level1}Numerical validation of extended model} 

The cumulative distribution expressed through Eq.~(8) or even its moderately simplified version in Eq.~(10) still fails to enable the direct estimation of activation parameters in the absence of \emph{a priory} knowledge of the distribution, \(\rho(G_{0},\nu_{0},v^{\ast})\). However, its effect on the overall cumulative distribution can be assessed by employing a simple form, which is easily verifiable through the Monte Carlo scheme presented earlier in Sec.~2. The simplest way to introduce statistical variations in the activation parameters is by assuming their distributions to be uniform around the corresponding mean values. We further assume that all the event types have the same attempt frequency, implying that the probability distribution of parameters only depends on the barrier-intercept, \(G_{0}\), and the activation volume, \(v^{\ast}\). Thus, \(G_{0}\) and \(v^{\ast}\) are randomly sampled from the ranges \(\lbrack\left\langle G_{0} \right\rangle - \mathrm{\Delta}G_{0},\ \left\langle G_{0} \right\rangle + \mathrm{\Delta}G_{0}\rbrack\) and
\({\lbrack\left\langle v^{\ast} \right\rangle - \mathrm{\Delta}v}^{\ast},\left\langle v^{\ast} \right\rangle + \mathrm{\Delta}v^{\ast}\rbrack\), respectively, and \(\rho(G_{0},v^{\ast}) = \frac{1}{4\mathrm{\Delta}G_{0}{\mathrm{\Delta}v}^{\ast}}\). Then, the cumulative probability is extracted from Eq.~(10) as,

\begin{eqnarray}
	\Phi(\sigma) &=& \frac{1}{4\Delta G_{0}\Delta v^{\ast}} \int_{\langle v^{\ast} \rangle - \Delta v^{\ast}}^{\langle v^{\ast} \rangle + \Delta v^{\ast}} \int_{\langle G_{0} \rangle - \Delta G_{0}}^{\langle G_{0} \rangle + \Delta G_{0}} \left[1 - \exp\left\{-\frac{\eta kT}{\dot{\sigma}v^{\ast}}\exp\left(\frac{\sigma v^{\ast}}{kT}\right)\right\}\right] \, dG_{0} \, dv^{\ast}
\end{eqnarray}

Equation~(12) can be validated by modifying the Monte Carlo method described in Sec.~2. At the beginning of each iteration, instead of using the fixed values of \(G_{0}\) and \(v^{\ast}\), we pick them randomly from a narrow range around the central values, \(\left\langle G_{0} \right\rangle\) and \(\left\langle v^{\ast} \right\rangle\). Figure~2 displays the cumulative distributions for \(\left\langle G_{0} \right\rangle\) = 1 with \(\mathrm{\Delta}G_{0}\) as 5\% of \(\left\langle G_{0} \right\rangle\). Three different central values of the activation volume, \(\left\langle v^{\ast} \right\rangle\) = 0.8, 1.0, and 1.2, have been used with variations of $\pm$5\%, $\pm$10\%, and $\pm$20\% for each. Similarly, the plots given in Fig.~3 correspond to $\pm$5\%, $\pm$10\%, and $\pm$20\% variation around \(\left\langle G_{0} \right\rangle\) with central values of \(\left\langle G_{0} \right\rangle\) = 0.8, 1.0, and 1.2, while the variation around \(\left\langle v^{\ast} \right\rangle\) is fixed at 5\% of \(\left\langle v^{\ast} \right\rangle\) = 1.\\

\begin{figure*}
	\centerline{\includegraphics*[width=18cm, angle=0]{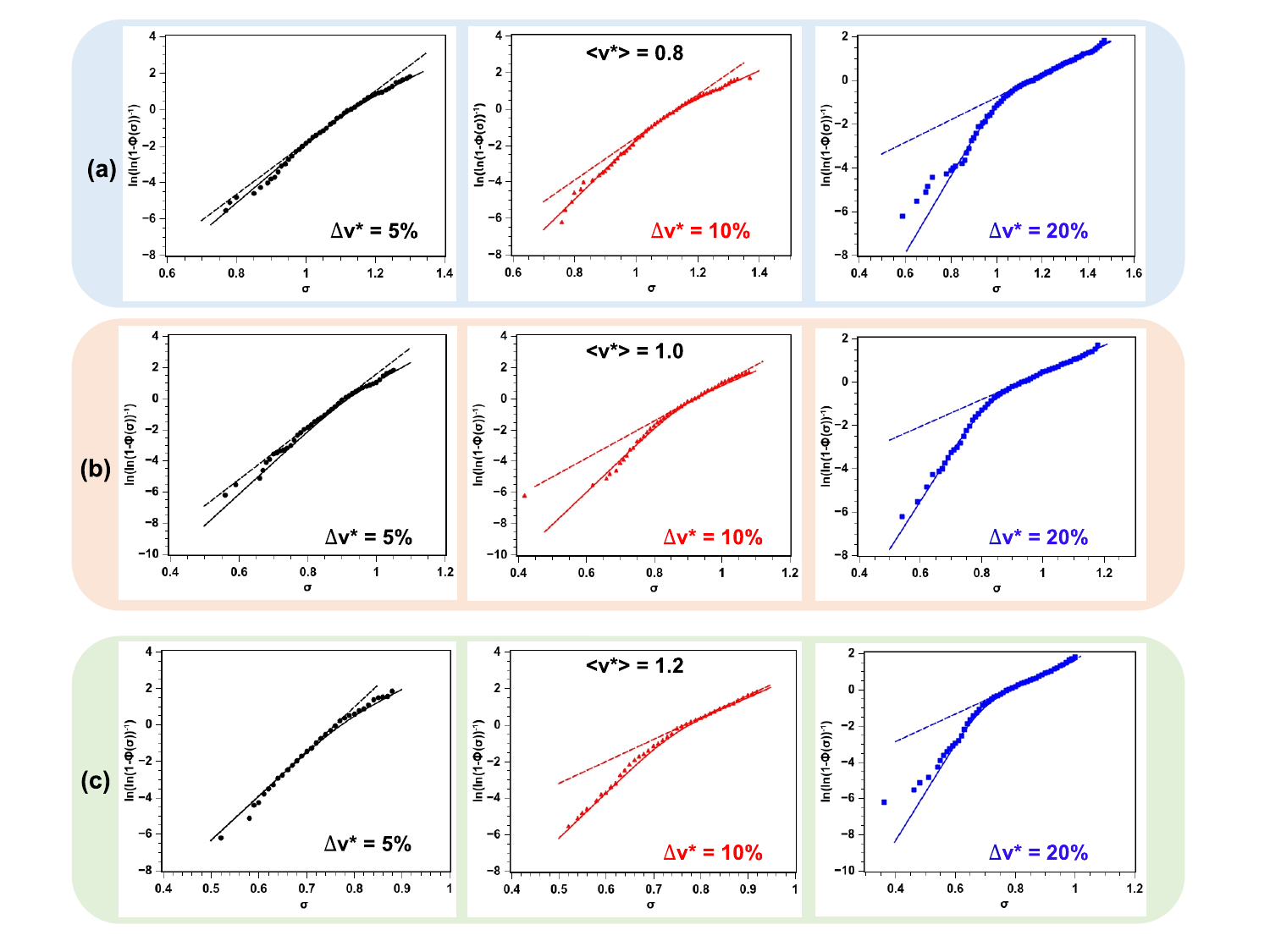}}
	\caption{The cumulative distributions ($\Phi$) obtained from the modified Monte Carlo scheme (scattered symbols) with \(\left\langle G_{0} \right\rangle\) = 1 and \(\mathrm{\Delta}G_{0}\) as 5\% of \(\left\langle G_{0} \right\rangle\). Three different central values of the activation volume, (a) \(\left\langle v^{\ast} \right\rangle\) = 0.8, (b) \(\left\langle v^{\ast} \right\rangle\) = 1.0, and (c) \(\left\langle v^{\ast} \right\rangle\) = 1.2, have been used with variations of $\pm$5\%, $\pm$10\%, and $\pm$20\%. The dashed lines are linear fits to the Monte Carlo results (scattered symbols), whereas the solid curves present the distributions obtained from Eq.~(12).}
\end{figure*}

Figures~2 and 3 display the cumulative distributions as double-logarithm of the reciprocal of survival probability, \(1 - \Phi(\sigma)\), similar to the representation in Fig.~1. We may recall that this double-logarithmic representation used in Fig.~1 and numerous earlier studies reported is inspired by the form of Eq.~(2) involving a double-exponential function, which leads to the linear behavior as verified in Sec. 2. Nevertheless, Eq.~(2) was derived assuming fixed activation parameters. In contrast, the cumulative distribution, \(\Phi(\sigma)\), given in Eqs.~(10) and (12) are observed to be the expectation of the distribution function, \(F(\sigma)\), given in Eq.~(2), over the ensemble of all the possible activation parameters.\\

\begin{figure*}
	\centerline{\includegraphics*[width=18cm, angle=0]{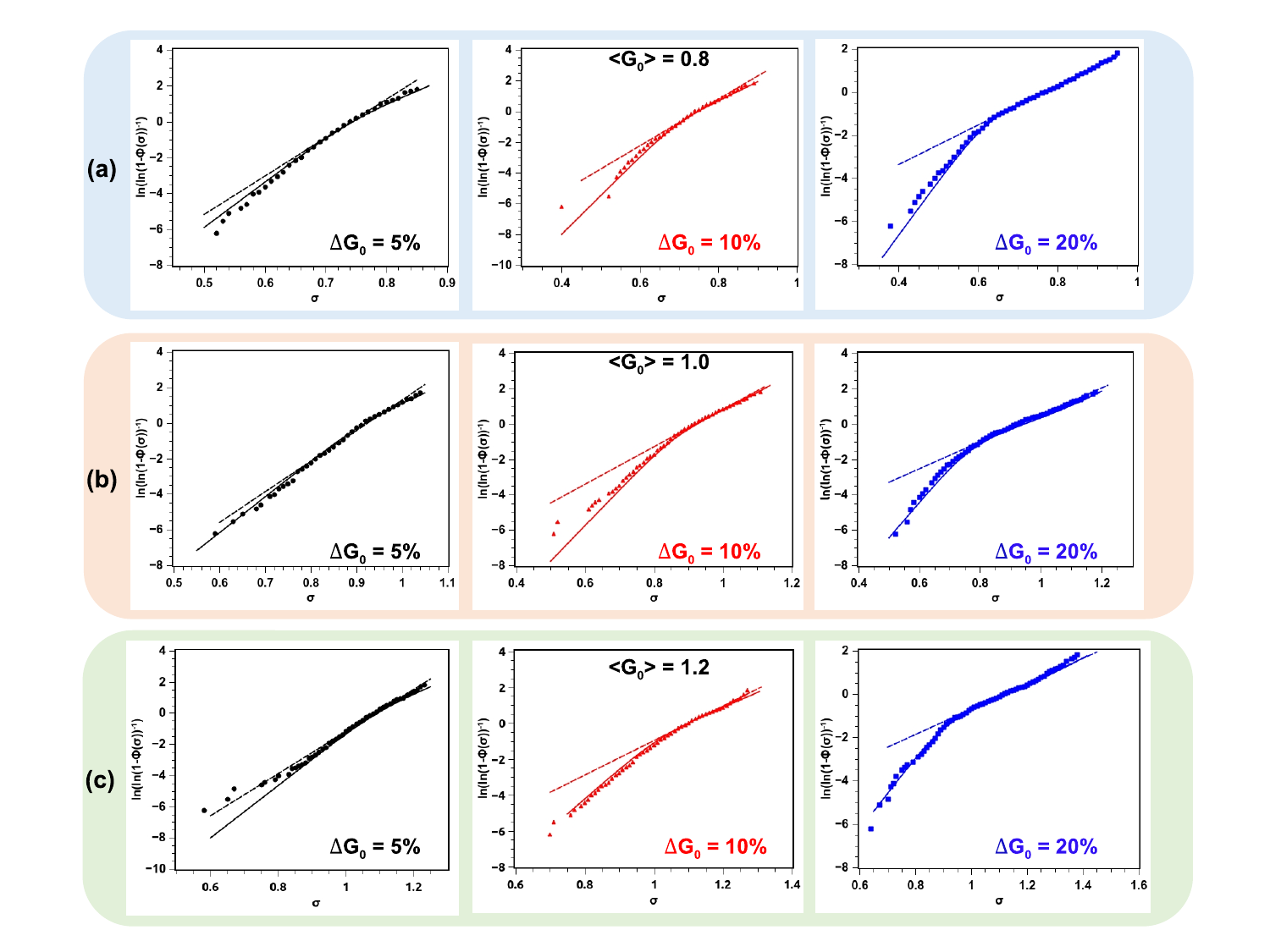}}
	\caption{ The cumulative distributions for \(\left\langle v^{\ast} \right\rangle\) = 1 with the variation around \(\left\langle v^{\ast} \right\rangle\) fixed at $\pm$5\%. Three different central values, (a) \(\left\langle G_{0} \right\rangle\) = 0.8 (b) \(\left\langle G_{0} \right\rangle\) = 1.0, and (c) \(\left\langle G_{0} \right\rangle\) = 1.2 have been considered with variations of $\pm$5\%, $\pm$10\%, and $\pm$20\%.  The curves and scattered symbols have the same representations as in Fig.~2. }
\end{figure*}

Unlike Eq.~(2), Eq.~(12) renders no reason to expect a linear behavior with double-logarithmic scaling, and Figs.~2 and 3 clearly reflect the non-linearity introduced by allowing the activation parameters to have small variance around their central values. The scattered symbols in Figs.~2 and 3 illustrate the results of the Monte Carlo calculations. In particular, we find that for a wider range over which the activation barrier, \(G_{0}\), and activation volume, \(v^{\ast}\), are allowed to vary, the curvature of the plot increases. Furthermore, the cumulative distributions are independently computed numerically using Eq.~(12) and are also displayed as solid curves. It can be seen that the Monte Carlo results are in excellent agreement with the analytically derived expression of Eq.~(12).

\section{\label{sec:level1}Physical interpretation of parametric variations}

In Sec.~2, we observed the validation of the original SL model using a simple Monte Carlo scheme, whereas its extended version is validated in Sec.~4. Here, it is pertinent to discuss whether the idea of parametric fluctuations, as introduced in Sec.~3, is merely a mathematical construct or founded on the physical nature of transition pathways. In the case of repeated measurements of an activated process under similar experimental conditions, one of the two scenarios is possible, which are schematically depicted in Fig.~4. In the first case (Fig.~4(a)), each instance of the process involves a transition from an initial state, $S_i$, to the final state, $S_f$, along a minimum energy path over the saddle point on the potential landscape. Accordingly, each measurement strictly corresponds to the same escape rate ($k$) from the initial state. In the second scenario, multiple transition pathways with various rates, $k_i$, may exist for the same initial state. As illustrated in Fig.~4(b), the transition can take the system from its initial state to one of the multiple final states along various transition pathways. For instance, we can consider the phenomenon of dislocation loop nucleation as the transition event. During two measurements of the critical load, loop nucleation can occur on one of the two or more available slip planes with different activation parameters. Similarly, defect nucleation can occur at more than one site under the same loading rate during a tensile or compressive deformation. Alternatively, a realistic potential landscape may be complex enough such that multiple pathways, each with its own saddle state and transition rate (illustrated as $k_3$, $k_4$, and $k_5$ in Fig.~4), can connect the initial state to the same final state. Several factors, like short-range compositional heterogeneity, variation in surface roughness, presence of oxide film, presence of defects like grain boundaries and precipitates, etc., can lead to the existence of multiple routes of transition. The average rate of escape from the initial state, $S_i$, would be the sum of individual rates corresponding to various transition paths. \\

\begin{figure}
	\centerline{\includegraphics*[width=18cm, angle=0]{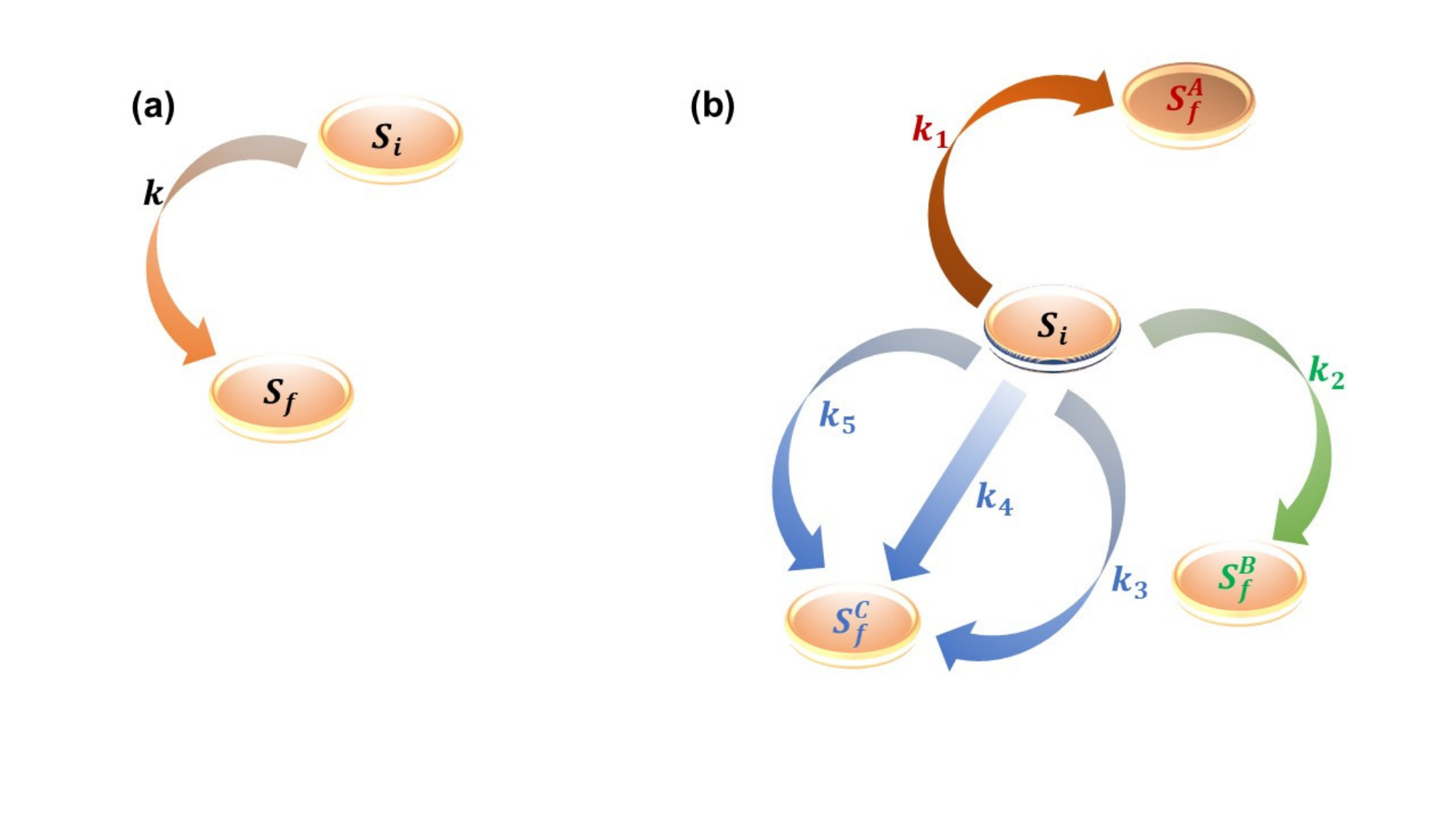}}
	\caption{Schematic representation of two possible scenarios for repeated measurements of an activated process under similar experimental conditions. (a) Each instance of the process involves a transition from an initial state, $S_i$, to a final state, $S_f$, along a minimum energy path, resulting in the same escape rate ($k$) from the initial state. (b) The transition can occur along multiple pathways with different rates ($k_i$), leading to various final states from the same initial state.}
\end{figure}

For a complex process like incipient plasticity, the idea picturized in Fig.~4(a) appears to fail to capture the underlying physical mechanism. Instead, the more generic notion of multiple escape pathways shown in Fig.~4(b), which is the basis of the extended model presented in Sec.~3, has been an intrinsic part of methods like kinetic Monte Carlo \cite{andersen2019practical} and accelerated molecular dynamics \cite{perez2015parallel}. Recent studies suggest that including more than one event type can indeed affect the statistical distribution of the critical load. Tan \textit{et al.} \cite{tan2023statistical} have shown that the generalized stacking fault energy in a complex concentrated alloy may be modeled as smoothly transitioning between two values depending on the size of the nucleated dislocation loop. This approximation can capture the effect of spatial compositional fluctuation and provide a better fit to the cumulative distribution. Similarly, Nisany and Mordehai \cite{nisany2022multiple} have theoretically analyzed the situation of multiple dislocation nucleation sites. Their MD simulations examined the nucleation of dislocation loops in nanowires with a rhombic cross-section, thereby providing acute and obtuse sites for possible heterogeneous nucleation. \\

It is worthwhile to recall that numerous studies have extensively resorted to Eq.~(3) resulting from the rigid-parameter SL model for determining the activation volumes. Despite the feasibility of multiple escape pathways as envisaged in Sec.~4 and depicted in Fig.~4(b), one can still attempt to fit the rigid-parameter model to the experimental measurements. This approach may be argued to be valid under the presumption that the activation parameters of multiple possible pathways exhibit sufficiently narrow statistical dispersion around their central values, such that Eq.~(3) becomes a close approximation of Eq.~(10). \emph{Prima facie}, this assumption seems reasonable. In the context of incipient plasticity, even if the repeated measurement of critical load emerges from various events, all such events are expected to be of a similar fundamental nature. Thus, the activation parameters cannot be presumed to vary drastically from one measurement to the next. Accordingly, one may be inclined to consider their estimates yielded by Eq.~(3) to be reasonably good representatives of their central values. In the subsequent section, we test this premise by directly applying the rigid-parameter SL model (Eq.~(3)) to the Monte Carlo results shown in Fig.~2.

\section{\label{sec:level2}Implications of parametric variations} 

The non-linear trend witnessed in Figs.~2 and 3 is not limited to the Monte Carlo calculations and is often observed in the experimental studies as well. In the original work by Schuh and Lund \cite{schuh2004}, from which the statistical model originated, the authors acknowledged that the tails of the cumulative distribution did not fit well with the analytical expression, and the linear fit was obtained for the region excluding the tails (Figs.~5 and 6 in \cite{schuh2004}). Curvatures in the experimentally obtained double-logarithmic plots were observed not only in their subsequent studies (Fig.~2 in \cite{schuh2005} and Fig.~5 in \cite{mason2006}) but several other experimental and simulation studies have shown that the analytical result rarely fits to the tails of the measured distribution. Non-linear double-logarithmic plots akin to the Monte Carlo and analytical results of Figs.~2 and 3 can also be seen in the reported studies on high-entropy alloys (\textit{e.g.}, Fig.~5 in \cite{zhu2013}) and metallic glasses \cite{tonnies2015rate,limbach2017serrated}. \\

In all the reported studies, the issue of non-linearity in the double-logarithmic plots is typically bypassed by selecting a convenient part of the whole distribution for fitting. To observe whether this approach can provide an acceptable estimate of the activation parameters, we select seemingly linear regions in the plots of Figs.~2 and 3 and attempt linear fits (dashed lines) based on the rigid-parameter approach. The resulting activation parameters computed based on Eq.~(3) are listed in Table 1, which reveals an astonishing trend. We find that the activation volumes, $v^*_{\text{lin}}$, measured in this manner with the brute-force assumption of linearity, are much smaller than the corresponding central values, \(\left\langle v^{\ast} \right\rangle\), used in the Monte Carlo calculations. Furthermore, the measured values drop even far below the entire range of \(v^{\ast}\). For instance, consider the case of \(\left\langle v^{\ast} \right\rangle\) = 0.8 and a variation of $\pm$20\% around this mean value, thus allowing \(v^{\ast}\) to vary from 0.64 to 0.96. For $\pm$5\% variation in \(G_{0}\) around \(\left\langle G_{0} \right\rangle\) = 1, the activation volume obtained using the linear fit is,  $v^*_{\text{lin}} = 0.26$, which is only a small fraction of the actual central value and much smaller than even the lower bound of 0.64. Hence, assuming the applicability of Eq.~(3) is observed to cause severe underestimation of the activation volume even for small variances of the activation parameters. A similar trend is found in Table 2 for the values of $G^0_{\text{lin}}$ estimated on the basis of Eq.~(3). Similar to the case of actual \textit{vs}. measured activation volumes, $G^0_{\text{lin}}$ can turn out to be significantly smaller than \(\left\langle G_{0} \right\rangle\) used in the Monte Carlo calculations.\\

\begin{table}[t]
	\caption{Activation volumes obtained from linear fits to the results of modified Monte Carlo calculations. Results are shown for $\langle G_{0} \rangle = 1$ with $\Delta G_{0}$ as 5\% of $\langle G_{0} \rangle$ and various extents of dispersion, $\pm\Delta v^{\ast}$, around the central values, $\langle v^{\ast} \rangle$.}
	\label{tab:mytable1}
	\begin{ruledtabular}
		\begin{tabular}{cccccc}
			$\langle v^{\ast} \rangle$ & $\Delta v^{\ast} = 0\%$ & $\Delta v^{\ast} = 5\%$ & $\Delta v^{\ast} = 10\%$ & $\Delta v^{\ast} = 20\%$ \\
			\hline
			0.8 & 0.85 & 0.71 & 0.58 & 0.26 \\
			1.0 & 1.11 & 0.84 & 0.59 & 0.31 \\
			1.2 & 1.30 & 1.21 & 0.60 & 0.38 \\
		\end{tabular}
	\end{ruledtabular}
\end{table}

Here, it is worth pointing out that a number of studies, mostly using nano-indentation experiments and the statistical model, have reported activation volumes of the order of atomic volumes, which are too small to be expected in dislocation-mediated plasticity. As the direct \textit{in situ} observation of the underlying mechanism of incipient plasticity has not been possible, some non-trivial mechanisms have been proposed to explain the ultra-small activation volumes. One expressed possibility pertains to the nucleation of numerous point defects beneath a nano-indenter tip \cite{schuh2004,golovin2002,farber1998mechanisms}. Another possibility raised by the authors was the nucleation of extremely small sub-equilibrium loops. A few other possible mechanisms were proposed in a subsequent study by Schuh
\textit{et al.} \cite{schuh2005}. One of them is the dislocation nucleation at pre-existing vacancy or vacancy-cluster. Stress concentration at surface asperities was also conjectured as a possible cause of nucleation with low activation volume. The authors also included the possibility of source activation through pre-existing line defects. Furthermore, in the study by Mason \textit{et al.} \cite{mason2006}, the mechanisms involving the migration of point defects were deemed untenable. Instead, surface asperities and heterogeneous nucleation in a supersaturated vacancy cloud were reiterated as plausible routes for the pop-in events. Zhu \textit{et al.} \cite{zhu2013} estimated the activation volume with first pop-in statistics in fcc high-entropy alloys and found it to be around three atomic volumes. They proposed that the incipient plasticity is governed by the vacancy-mediated heterogeneous nucleation of dislocation loops with a critical size of only \(\sim3\) atoms. A similar activation volume was reported for the bcc high-entropy alloy \cite{ye2017dislocation}, and the authors suggested the mechanism of cooperative migration of a few atoms to be at play.\\

\begin{table}[h!]
	\caption{Barrier intercepts ($G^0_{\text{lin}}$) calculated from the linear fits to Monte Carlo results obtained with $\langle v^{\ast} \rangle = 1$ and $\Delta v^{\ast}$ as 5\% of $\langle v^{\ast} \rangle$. Results for multiple values of dispersion, $\pm\Delta G_{0}$, around the central values, $\langle G_{0} \rangle$, are presented here.}
	\label{tab:mytable2}
	\centering
	\begin{ruledtabular}
		\begin{tabular}{ccccc}
			$\langle G_{0} \rangle$ & $\Delta G_{0} = 0\%$ & $\Delta G_{0} = 5\%$ & $\Delta G_{0} = 10\%$ & $\Delta G_{0} = 20\%$ \\
			\hline
			0.8 & 0.88 & 0.85 & 0.60 & 0.36 \\
			1.0 & 1.11 & 0.86 & 0.53 & 0.38 \\
			1.2 & 1.02 & 0.81 & 0.57 & 0.35 \\
		\end{tabular}
	\end{ruledtabular}
\end{table}

The findings reflected in Tables~1 and 2 indicate that the brute-force approach of fitting the rigid-parameter model (Eq.~3) to a process incorporating statistical fluctuations in the activation parameters may lead to severe underestimation, even if the fluctuations are small. It suggests an alternative perspective explaining the ultra-small activation volumes reported in the experimental and simulation studies without resorting to various non-trivial mechanisms. As discussed in Sec.~3 and 5, the activation parameters are likely to exhibit some variation during repeated measurements of the critical loads for various reasons. It is evident from the Tables that fluctuations of only a few percent around the central value are enough to suppress the measured activation volume grossly. It hints at the possibility that the statistically estimated activation volumes of the order of atomic volume could merely be the outcome of rigidly fixing the activation parameters in the statistical model, a premise that is inapt against the backdrop of physical plausibility. The actual activation volumes can be larger than those obtained from the linear fits owing to small parametric variations. \\

\section{\label{sec:level1}Deformation simulations of Iron nanopillars} 

So far, we have observed that the rigid-parameter model can be extended to incorporate the effect of fluctuations. Although the analytical results and Monte-Carlo calculations for parametric fluctuations reveal underestimated activation volume yielded by the rigid-parameter model, observation of such effect in a physical deformation process is an intriguing possibility. To further explore this aspect, we perform MD simulations of uniaxial tension and compression tests on virtual Fe nanopillars. The purpose of this investigation is twofold; on the one hand, the tension-compression asymmetry seen in bcc metals can be perceived from the standpoint of statistical analysis, and on the other hand, it offers a clean and ideal system having the conventional modalities of deformation without involving exotic mechanisms that may render exceptionally small activation volumes.

\subsection{\label{sec:level2}Simulation Scheme} 

The interatomic force-field employed in the simulations is derived from the embedded atom method (EAM) potential developed by Mendelev and co-workers \cite{mendelev}. In addition to replicating the basic material properties, this potential correctly reproduces the non-degenerate core structure of screw dislocations and displaced twin boundaries in agreement with the DFT calculations \cite{chaussidon2006glide}.\\

\begin{figure}
	\centerline{\includegraphics*[width=14cm, angle=0]{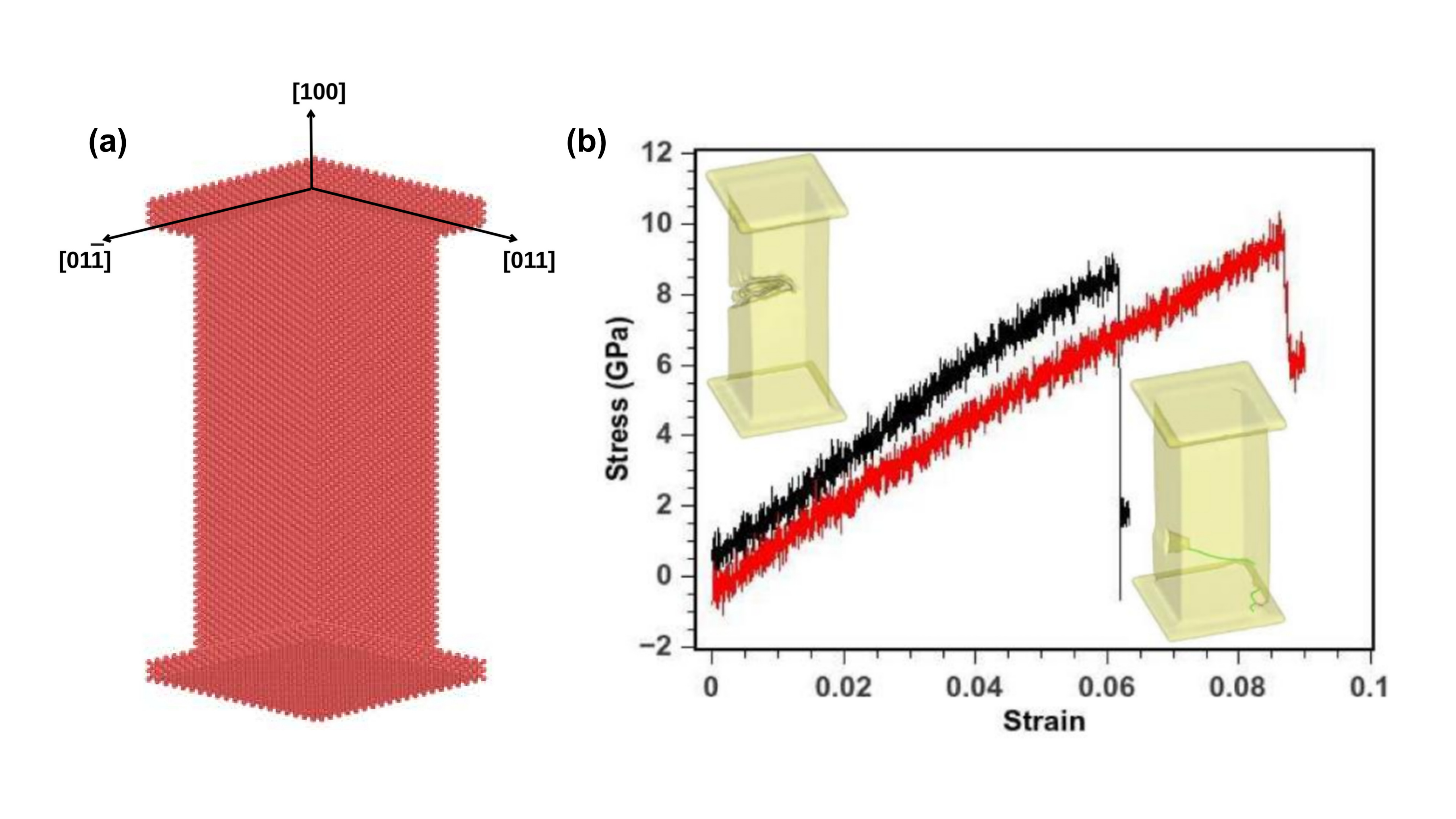}}
	\caption{(a) The simulated bcc-Fe nanopillar with crystal directions along the axis and normal to the surfaces. (b) The engineering stress on the Fe nanopillar plotted against both tensile and compressive strains. The black plot represents the tensile loading while the red plot corresponds to the compressive deformation. The accompanying simulation snapshots show the formation of twinned region in tension and the screw dislocation in compression. }
\end{figure}

The 15.4 nm long nanopillar has a 5.8 nm $\times$ 5.8 nm square cross-section between two larger indenter plates, as shown in Fig.~5(a). The \textless100\textgreater-oriented pillar has low-energy \{110\}-surfaces, which prevent some spurious deformation effects associated with the high-energy surfaces \cite{healy2014molecular}. Uniaxial deformation is applied at a strain rate of $10^7 \, \text{s}^{-1}$ and 300 K temperature with a time-step of 2 fs for the velocity-Verlet integrator. Two sets of simulations with random velocity initializations for tensile and compressive loading have been carried out, with 80 simulations in each set. During the deformation, the forces on the indenter plates are recorded to measure the yield stresses on the nanopillar.\\

In addition to the MD simulations, the yielding mechanisms have been examined through the nudged-elastic band calculations. To this end, we extract the instantaneous structures just before and after the yield point. These structures are axially strained until the desired stress is obtained and are subsequently used to initialize the chain of replicas. This chain evolves iteratively using the modified inertial relaxation method with the climbing replica scheme \cite{climbing,FIRE_algorithm}. The NEB computations reveal the pathways of the elastic-to-plastic transitions in both tensile and compressive deformations. All the computations reported in this work have been carried out using the LAMMPS MD code \cite{2022lammps}. The OVITO program with dislocation extraction analyzer \cite{stukowski2009,DXA} aids the visualization and analysis of crystalline defects.

\subsection{\label{sec:level2}Simulations results and discussion}

Figure~5(b) exhibits the typical stress \textit{vs}. strain plots of the nanopillars under tensile and compressive loadings. The transition from elastic to plastic deformation is easily identified as the sudden drop in measured load at the yield point. As observed in earlier studies \cite{dutta2017,healy2014molecular,shi2016competing}, the plasticity of tensile deformation is governed by the formation and growth of a twinned region. In contrast, the compressive yielding is controlled by the nucleation and glide of screw dislocations. The snapshots of simulation cells accompanying the stress-strain plots in Fig.~5(b) reveal these two distinct deformation mechanisms.\\

\begin{figure}[t!]
	\centerline{\includegraphics*[width=7cm, angle=0]{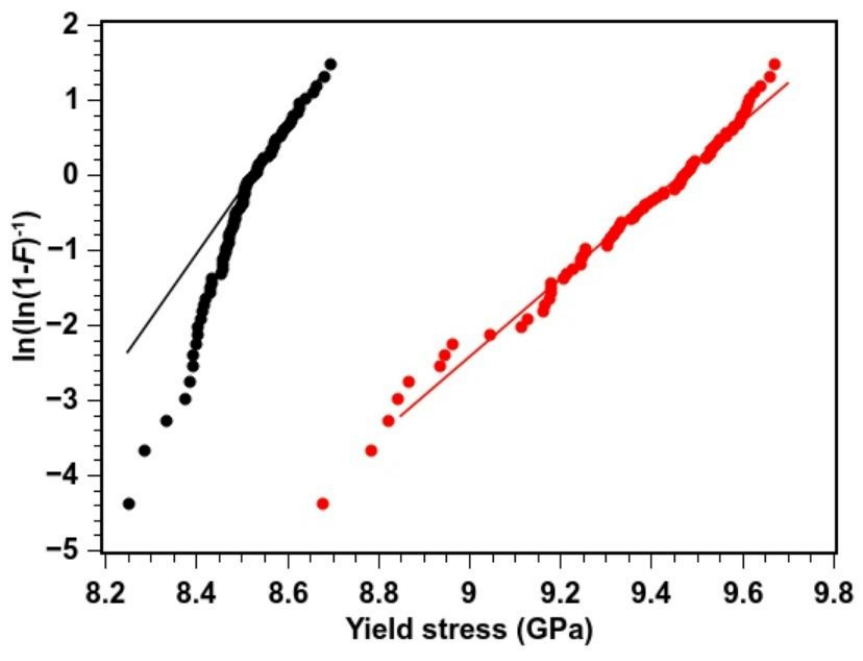}}
	\caption{The cumulative probabilities (\emph{F}) of the yield stresses measured in the MD simulations of Fe nanopillars. The black circles are the results from tensile loading whereas the red ones represent compressive deformation. The solid lines are the linear fits to measured distributions.}
\end{figure}

As specified in Sec.~7.1, multiple randomized simulations have been performed for both types of deformation. Figure~6 shows the reciprocals of survival probabilities ($1-F$) plotted with a double-logarithmic scale similar to that in Fig.~1. We observe that the linear trend is valid only for a part of the plot with tails deviating from the trend predicted in Eq.~(3), though the results of compressive loading provide a better fit over a broader range than their tensile counterparts. This tail-effect is consistent with the results observed in numerous nano-indentation experiments. As discussed earlier, the tails are considered less reliable part of the measured distribution and often omitted while presenting the results.\\

By fitting the rigid-parameter model to the distribution shown in Fig.~6, the activation volumes for tensile and compressive yielding of the nanopillar are obtained as 3.09 and 1.83 (atomic volume), respectively. It must be pointed out that in Eq. (3), we directly use the normal stress, $\sigma$, instead of its resolved shear component, $\tau$. As a result, the activation volume referred to in this work is assumed to be defined as, $v^* = -\frac{\partial G}{\partial \sigma}$, which is typically expected to be $\sim$2-3 times smaller than the shear activation volume, $-\frac{\partial G}{\partial \tau}$, on account of the Schmid-factor. We observe that the activation volume obtained for the compressive deformation is substantially smaller than that for the tensile deformation. This finding is apparently in consonance with our intuitive understanding, as we expect the heterogeneous nucleation of a full dislocation loop to involve fewer atoms than a twin nucleus involving multiple atomic layers and a larger volume. However, even though the statistical analysis reveals the signature of tension-compression asymmetry, the perplexingly small activation volumes still warrant further analysis.\\

NEB calculations provide a detailed view of the mechanisms involved in yielding under the tensile and compressive loads. As observed in Fig.~7(a), the twin fault nucleation occurs at the surface, and the twinned region grows gradually in tensile deformation. Similarly, in compressive deformation (Fig.~7(b)), the dislocation loop is seen to nucleate at the corner, which soon re-orients to a screw character due to the drastic difference in mobilities of edge and screw components \cite{dutta2017}. The transition pathways for both types of deformation indicate that, as expected, the processes are mediated simply by the planar slips. Furthermore, the mechanisms involve the conventional nucleation of full or twinning partial dislocations and do not involve any non-trivial phenomenon that can justify the unusually small activation volumes estimated from the rigid-parameter model. These observations indicate that the original SL method can exhibit such activation volumes even with the conventional modalities of deformation. It is also pertinent to point out that on account of significantly lower rates of deformation, the yield stresses in experiments are much smaller than those obtained from MD simulations. As the activation volumes of events like dislocation nucleation and depinning typically elevate rapidly with reduction in the critical stress \cite{aubry2011energy,jennings2013modeling,faisal2019exponent,ryu2011entropic,niewczas2015thermally}, we can expect the activation volumes in experimental studies to be even larger than the simulations, which renders the earlier reported results even more confounding.\\

\begin{figure}
	\centerline{\includegraphics*[width=15cm, angle=0]{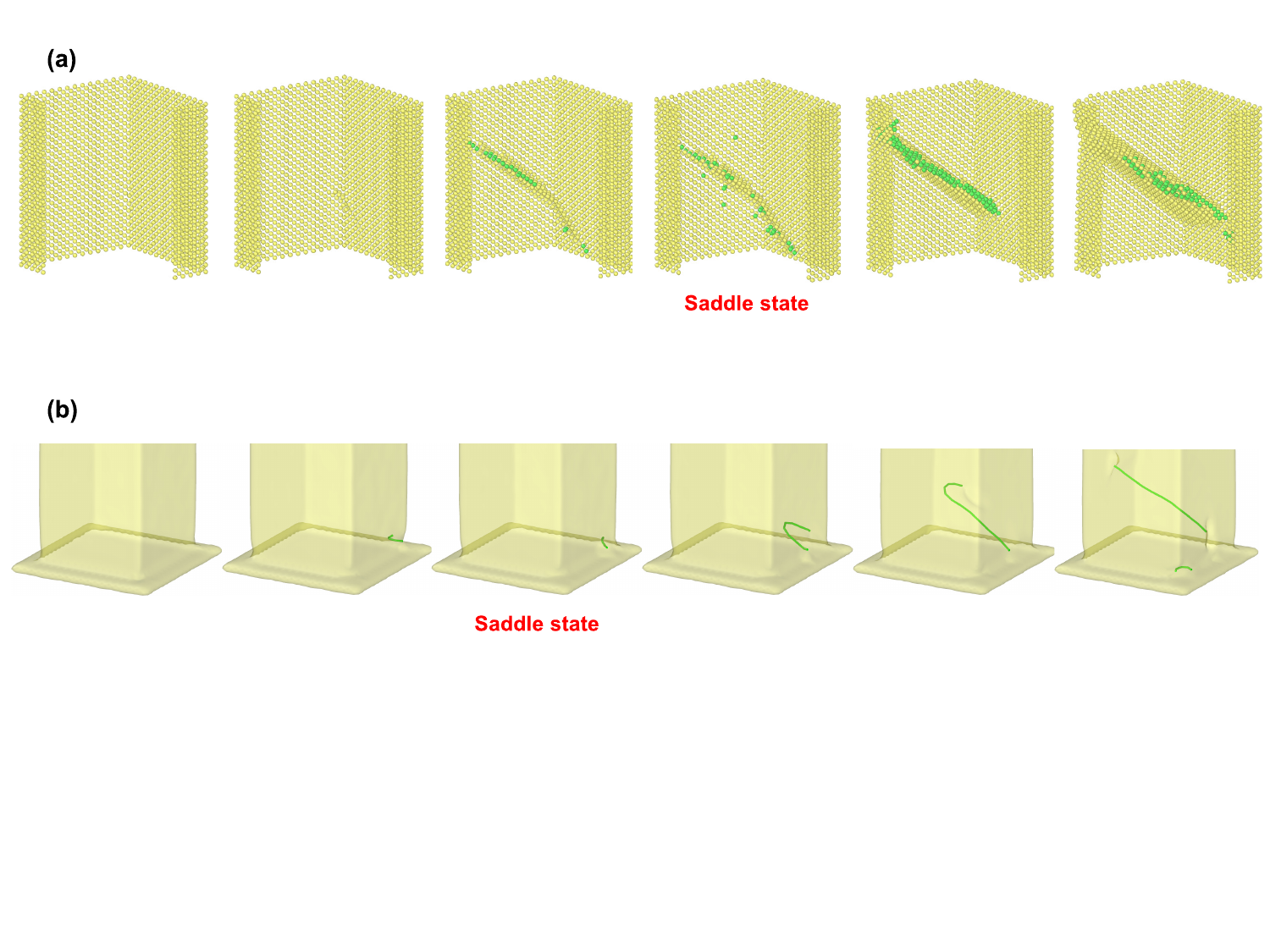}}
	\vspace{-90pt} 
	\caption{Pathways of incipient plasticity of Fe nanopillars as revealed by the NEB computations for (a) tensile and (b) compressive modes of deformation.  }
\end{figure}

Section~6 presented a brief survey of various mechanisms, which have been conjectured to explain the unusually small activation volumes. Nevertheless, a bird's-eye view of the reported results presents a perplexing observation. The statistical model has been applied to a variety of experimental results obtained for pure metals \cite{wu2014incipient,mason2006}, ceramics \cite{stich2022room}, high-entropy alloys \cite{zhu2013,ye2017dislocation,gan2021interstitial,zhao2021bimodality,mridha2019activation}, and even metallic glasses \cite{tonnies2015rate,limbach2017serrated}. In each of these studies, the activation volumes are invariably found to be of atomic order, irrespective of the choice of material and mechanism of deformation. The same trend is maintained in the reported MD simulations of Mo nanoparticles \cite{chachamovitz} and the Fe nanopillars, as seen in the present work. As a matter of fact, none of the reported studies has ever estimated a statistically computed activation volume that is in consonance with the conventional understanding of the  mechanisms of incipient plasticity.\\

Interestingly, activation volumes of the order of atomic volume have been inconsistent with the other measurement methods. For instance, Brandl \textit{et al.} \cite{brandl2022} studied the temperature-controlled nano-indentation of bcc chromium and, through a different method, estimated the activation volume to be one or two orders of magnitude larger than those typically obtained from the statistical model. Surprisingly, this estimation was for the process of kink nucleation, which is expected to involve only a few atoms. Similarly, direct atomistic computations have estimated the activation volumes of heterogeneous nucleation of dislocation loops, which are one to two orders of magnitude larger than those calculated from the prevailing statistical method \cite{zhang2022atomistic,du2016mechanism}. Given that the ultra-small activation volumes of incipient plasticity are estimated in a variety of crystalline and amorphous materials, it appears that this trend can rather be a ubiquitous and inherent feature of the rigid-parameter model than the outcome of special mechanisms.\\

Even if one ignores the loose interpretation of activation volume in terms of the number of atoms involved in a critical event, the straightforward definition still indicates that a very small activation volume implies extreme insensitivity of the activation barrier towards changes in the applied load. It is indeed counterintuitive, for shear-driven mechanisms like dislocation nucleation in crystalline solids or STZ activation in glassy alloys are expected to be load-sensitive. Extension of the SL model (Sec.~3) along with the Monte Carlo and MD computations suggest that the unusually small activation volumes, as ubiquitously reported in the earlier studies, can rather be an inherent feature of the rigid-parameter approach than the consequence of special nucleation mechanisms. Clearly, there is a need of having a closer look at this issue, for the findings and arguments presented here put forward a strong case in favor of reframing the statistical approach in the backdrop of parametric variations.\\

\section{\label{sec:level1}Conclusions}

In a nutshell, the present work closely examines the strategy of estimating the activation parameters of critical events associated with incipient plasticity by analyzing the cumulative distribution of critical loads. It is shown that the existing statistical method remains valid only under the assumption of rigidly fixed activation parameters, where the variance in the measured critical loads is solely attributed to thermal activation. In a more physically plausible scenario, the activation parameters are expected to fluctuate during the sequence of measurement. We extend and generalize the statistical model to allow for such statistical parametric fluctuations and find that even a small variance in the activation parameters can significantly alter the distribution of critical loads compared to the conventional rigid model. In particular, fitting the rigid-parameter model to the distribution obtained with parametric fluctuations can potentially produce the misleading result of extremely small activation volumes. This may explain why the experiments and simulations have ubiquitously reported activation volumes comparable to atomic volumes.\\

While the present study underlines the effect of not incorporating the parametric fluctuations within the constrained statistical model's purview, another issue remains open. A generalization of the rigid model to its unconstrained version shows that the cumulative distribution of critical loads depends on the distribution of activation parameters. The present study verifies this by considering a toy distribution, where the parameters are uniformly distributed over a narrow range. This assumption, though simple, captures two essential components of a distribution: a central value and a measure of dispersion. For a realistic activation mechanism of incipient plasticity, either in an experiment or simulation, we do not have access to the exact form of the parametric distribution. A possible workaround is to assume a reasonable distribution like a Gaussian or log-normal probability density and treat its mean and standard deviation as free parameters. Subsequently, fitting can be performed using the data from multiple sets of measurements at various temperatures and loading rates by tuning the distribution parameters. The outcome of this work can pave the way for further investigations in that direction.

\section*{\label{sec:level1}Data Availability}

Data will be made available on request.

\bibliography{apssamp}

\end{document}